\documentclass[a4paper,11pt]{article}

\usepackage{jheppub}

\usepackage{subfig}
\usepackage{amsmath}
\usepackage{amssymb}
\usepackage{latexsym}
\usepackage{wasysym}
\usepackage{pstricks}

\newcommand{\msbar}{\overline{\mbox{MS}}}

\newcommand{\ice}[1]{\relax}

\newcommand{\ed}{\end{document}}

\newcommand{\prd}{\partial}
\newcommand{\ep}{\epsilon}
\newcommand{\beq}{\begin{equation}}
\newcommand{\eeq}{\end{equation}}
\newcommand{\bea}{\begin{eqnarray}}
\newcommand{\eea}{\end{eqnarray}}

\newcommand{\ba}{\begin{array}}
\newcommand{\ea}{\end{array}}

\newcommand{\g}{\gamma}
\newcommand{\al}{\alpha}

\newcommand{\bc}{\begin{center}}
\newcommand{\ec}{\end{center}}

\newcommand{\re}[1]{(\ref{#1})}
\newcommand{\unl}[1]{\underline{#1}}

\catcode`\@=11
\def\slash{\mathpalette\make@slash}
\def\make@slash#1#2{\setbox\z@\hbox{$#1#2$}%
  \hbox to 0pt{\hss$#1/$\hss\kern-\wd0}\box0}
\catcode`\@=12 

\def\bbuildrel#1_#2^#3%
{\mathrel{\mathop{\kern 0pt#1}\limits_{#2}^{#3}}}

\newcommand{\sbz}{  }
\newcommand{\nnb}{\nonumber}
\newcommand{\as}{a_s}

\newcommand{\MSbar}{\ensuremath{\overline{\text{MS}}}\ }

\newcommand{\tnf}{T_f\,n_f }

\newcommand{\tnfB}{T^2_f\,n^2_f }

\newcommand{\tnfC}{T^3_f\,n^3_f }

\newcommand{\tnfD}{T^4_f\,n^4_f }

\newcommand{\cfB}{C_F^2 }
\newcommand{\cfAcaA}{C_F \, C_A }
\newcommand{ \cfAtrAnlA}{ C_F\, T_f\,n_f }

\newcommand{\cfC}{C_F^3 }

\newcommand{\cfAtrBnlB}{C_F \,  T_f^2\,n_f^2 }

\newcommand{\cfBtrAnlA}{C_F^2\, T_f\,n_f }

\newcommand{\cfAtrAnlAcaA}{C_F\, \tnf \, C_A }

\newcommand{\cfBcaA}{C_F^2\, C_A}

\newcommand{\cfAcaB}{C_F\, C_A^2}

\newcommand{\cfD}{C_F^4 }
\newcommand{\cfAtrCnlC}{C_F\, \tnfC }
\newcommand{\cfBtrBnlB}{C_F^2\,\tnfB}

\newcommand{\cfAtrBnlBcaA}{C_F\, \tnfB \,C_A}

\newcommand{\cfCtrAnlA}{C_F^3\,\tnf}
\newcommand{\cfBtrAnlAcaA}{C_F^2 \,C_A\,\tnf}

\newcommand{\cfAtrAnlAcaB}{C_F \,\tnf\,C_A^2}

\newcommand{\cfCcaA}{C_F^3 \,C_A}
\newcommand{\cfBcaB}{C_F^2 \,C_A^2}
\newcommand{\cfAcaC}{C_F \,C_A^3}

\newcommand{\dFFdRinvAnlA}{n_f\,\frac{d_F^{abcd}\,d_F^{abcd}}{d_R}}
\newcommand{\dFARinvA}{\frac{d_F^{abcd}\,d_A^{abcd}}{d_R}}

\newcommand{\cfE}{C_F^5 }

\newcommand{\cfBtrCnlC}{C_F^2\, \tnfC }

\newcommand{\cfCtrBnlB}{C_F^3\, \tnfB}

\newcommand{\cfDtrAnlA}{C_F^4\,\tnf}

\newcommand{\cfCtrAnlAcaA}{ C_F^3 \, \tnf\,C_A }

\newcommand{\cfBtrAnlAcaB}{ C_F^2\,\tnf \,C_A^2\,}

\newcommand{\cfDcaA}{C_F^4 \,C_A}

\newcommand{\cfCcaB}{C_F^3 \,C_A^2}

\newcommand{\cfBcaC}{C_F^2 \,C_A^3}

\newcommand{\cfAtrDnlD}{C_F\, \tnfD }

\newcommand{\cfAtrCnlCcaA}{C_F\, \tnfC \,C_A}

\newcommand{\cfBtrBnlBcaA}{C_F^2\,\tnfB \,C_A}

\newcommand{\cfAtrBnlBcaB}{C_F \,\tnfB \,C_A^2}

\newcommand{\cfAtrAnlAcaC}{C_F \,\tnf\, C_A^3}

\newcommand{\cfAcaD}{C_F \,C_A^4}

\newcommand{\dFFdRinvAnlAcfA}{ C_F\, n_f\,\frac{d_F^{abcd}\,d_F^{abcd}}{d_R} }  

\newcommand{\dFFdRinvAnlBtrA}{ T_f\, n_f^2\,\frac{d_F^{abcd}\,d_F^{abcd}}{d_R}}     

\newcommand{\dFFdRinvAnlAcaA}{ n_f\,C_A\, \frac{d_F^{abcd}\,d_F^{abcd}}{d_R}}

\newcommand{\dFARinvAcfA}{\frac{C_F\, d_F^{abcd}\,d_A^{abcd}}{d_R}}         
 
\newcommand{\dFARinvAnlAtrA}{\tnf\, \frac{d_F^{abcd}\,d_A^{abcd}}{d_R}}      
                       
\newcommand{\dFARinvAcaA}{C_A\,\frac{d_F^{abcd}\,d_A^{abcd}}{d_R}}

\newcommand{\dAARinvAtrA}{T_f\,\frac{d_A^{abcd}\,d_A^{abcd}}{d_R}}

\def\beq{\begin{equation}}
\def\eeq{\end{equation}}
\def\bea{\begin{eqnarray}}
\def\eea{\end{eqnarray}}
\def\bq{\begin{quote}}
\def\eq{\end{quote}}

\def\nnb{\nonumber}

\def\nnb{\nonumber}

\def\ba{\begin{array}}
\def\ea{\end{array}}

\newcommand{\mbib}{ 
\bibliographystyle{JHEP}
\bibliography{chet,chet_full,lit,gm5,rg,higgs_as5,dim_reg,JJ,QGRAF_EXP,LiteraturSM,%
steinhauser,baikov,asmirnov,smirnov,vladimirov,vermaseren,beta5,%
surguladze,laporta,gorishnii,tarasov,bierenbaum,kataev,czakon,kazakov,david_dirk_qQED,other_masters,%
sector_decom,lee,kotikov_before1995,remiddi_1997-2000,broadhurst,gracey,velizhanin,%
DIS,DIS2,DIS4,acat,LiteraturSM}
}

\title{
{
\boldmath
\!\!\!\!\!\!\!\!
Five-loop  fermion anomalous dimension 
\\
for a general gauge group from 
four-loop massless propagators  
}
}
\author[a]{P.~A.~Baikov,}
\author[b]{K. G. Chetyrkin, }
\author[b]{J.~H.~K\"uhn}

\affiliation[a]{
Skobeltsyn Institute of Nuclear Physics, Lomonosov Moscow State University, 
1(2), Leninskie gory, Moscow  119991, Russian Federation
        }        
\affiliation[b]{Institut f\"ur Theoretische Teilchenphysik, Karlsruhe
  Institute of Technology (KIT), Wolfgang-Gaede-Stra\ss{}e 1, 726128 Karlsruhe, Germany}

\emailAdd{baikov@theory.sinp.msu.ru}
\emailAdd{Konstantin.Chetyrkin@kit.edu}
\emailAdd{johann.kuehn@kit.edu}

\abstract{

We extend the ${\cal O}(\alpha_s^5)$ result of the analytic calculation of the quark mass anomalous
dimension  in pQCD \cite{Baikov:2014qja} to the case
of a generic gauge group. We present explict formulas which express the
relevant renormalization constants in terms of four-loop  massless propagators.
We also use our result to shed  new light on the old puzzle 
of the absence of even zetas in results of perturbative calculations
for  a  class of physical observables. 
}


\begin{document}

\maketitle

\section{Introduction \label{sec:intro}}

The  dependence of the quark masses on the   renormalization scale is governed by   
the quark mass anomalous dimension, defined  as
\begin{equation}
\mu^2\frac{d}{d\mu^2} {m}|{{}_{{g_0},
 m_0 }}
 = {m} \gamma_m(a_s) \equiv
-{m}\sum_{i\geq0}\gamma_{{i}}
\,
a_s^{i+1}
{},
\label{anom-mass-def}
\end{equation}
where $a_s = \alpha_s/\pi= g^2/(4\pi^2)$, $g$ is the renormalized
strong coupling constant and $\mu$ is the normalization scale in the
customarily used $\msbar$ renormalization scheme.

Up to and including four-loop level the anomalous dimension is known since 20 years
\mbox{\cite{Tarrach:1980up,Tarasov:1982gk,Larin:1993tq,Chetyrkin:1997dh_good,Vermaseren:1997fq}}.
Some time ago this result has been pushed to five-loop order, albeit for the
special case of an $SU(3)$ theory with $n_f$ species of quarks \cite{Baikov:2014pja,Baikov:2014qja}.
In the present paper the corresponding result will be given for an arbitrary
compact simple Lie group.

The quark mass anomalous dimension has important implications: the Higgs boson
decay rate into charm and bottom quarks, respectively, is proportional to the
square of the respective quark mass at the scale of $m_H$, which, for given
on-shell mass, depends on the fore-mentioned anomalous dimension 
(see, e.g., recent discussions in 
\cite{Baikov:2014pja,Baikov:2014qja,Baikov:2016tgj,Chetyrkin:2016uhw,Chetyrkin:2016pbz,Baikov2016}).  The
generalization of the result from $SU(3)$ to an arbitrary compact semi simple
Lie group, as presented in this paper, gives additional insight into the
structure of the result.

This paper is organized as follows: technical preliminaries will be presented
in Section 2. The next Section 3 briefly discusses master formulas for the two
relevant renormalization constants which define quark mass anomalous
dimension.  The main result, the generalization of 
the ${\cal O}(\alpha_s^5)$ expression for $\g_m$ 
from $SU(3)$ to an arbitrary
compact simple Lie group will be presented subsequently in Section 4, together
with the special case of QED. In Section 5 we will discuss in some detail the
structure of irrational contributions to the result for $\gamma_m$.  Our short
conclusions will be given in Section 6.

\section{Technical preliminaries}

The quark mass renormalization constant, 
$Z_{m}$,  is defined as the ratio of the bare and renormalized
quark masses, viz.  
\beq
Z_m = \frac{m_0}{m} = 1  +
\sum_{i,j}^{0<j\leq i}\left(Z_{{m}}\right)_{ij}
\frac{ \as^i   }{\epsilon^{{j}}}
\label{Zm}
{}.
\eeq
Within the
\MSbar\  scheme \cite{tHooft:1972fi,Bardeen:1978yd}
the coefficients $\left(Z_{{m}}\right)_{ij}$ are just numbers; 
$\epsilon \equiv 2 - D/2 $ and $D$ stands for 
the space-time dimension. Combining eqs.~(\ref{anom-mass-def},\ref{Zm}) and
using the RG-invariance of  $m_0$, one arrives at the following
formula for $\g_m$:
\beq
\gamma_m = \sum_{ i \ge 0}  (Z_m)_{i1} \,i\,  a_s^i
{}.
\eeq

One possibility to obtain  $Z_m$ is to use the  well-known relation (see, e.g. Section 2.2 of \cite{Chetyrkin:1996ia})
\beq
Z_m = Z_{\bar{\psi} \psi} /Z_2
\label{Zm:def}
{},
\eeq
where  $Z_{\bar{\psi}\psi}$ is
the renormalization constant of the 
quark mass operator\footnote{We use square  brackets to refer to completely UV renormalized quantities; 
for  simplicity we also do not write explicitly the \MSbar renormalization scale $\mu$.}   
\beq
[ \bar{\psi} \psi ] =  Z_{\bar{\psi} \psi}\,  {\bar{\psi} \psi} 
=  Z_{\bar{\psi} \psi}/Z_2 \,  {\bar{\psi_0} \psi_0}
\eeq
and $Z_2$ is  the quark field renormalization constant
\beq
\psi_0 = Z_2^{1/2} \psi
{}.
\eeq
The QED  Ward identity implies that $Z_2\equiv Z_V$, with $Z_V$ being the renormalization constant of the quark 
vector current: 
\beq
[ \bar{\psi}\g_\al \psi ] =  Z_V\,  {\bar{\psi} \g_\al\psi} 
=  Z_V/Z_2 \,  {\bar{\psi_0}\g_\al \psi_0} = {\bar{\psi_0}\g_\al \psi_0}
{}.
\eeq
To compute $Z_2$ and $Z_{\bar{\psi} \psi}$ one  starts from  the corresponding  bare vertex functions
(we assume that the only external momentum, $q$ ,is flowing in and out through the fermion legs) 
\beq 
 \g_\al\, \Gamma^V_B(\as^0,q^2) = \g_\al\,\left(1 + \delta \Gamma^V_B(\as^0,q^2)\right)
{}
\eeq 
and 
\beq 
\Gamma^S_B(\as^0,q^2) =  1 + \delta \Gamma^S_B(\as^0,q^2)
{}.
\eeq 
Requiring the finiteness of the renormalized versions of  both vertex functions 
\beq
[\Gamma^V](\as,q^2) = Z_V \, \Gamma_B(\as^0,q^2) = Z_V +  Z_V \, \delta \Gamma_B(\as^0,q^2)
\label{Gvector:def}
{},
\eeq
\beq
[\Gamma^S](\as,q^2) = Z_{\bar{\psi} \psi} \, \Gamma_B(\as^0,q^2) = Z_{\bar{\psi} \psi} +  Z_{\bar{\psi} \psi} \, \delta \Gamma^S_B(\as^0,q^2)
\label{Gscalar:def}
\eeq
we arrive 
at the following recursive equations  for $Z_2= 1 + \delta Z_2$ and 
$Z_{\bar{\psi} \psi} = 1 + \delta Z_{\bar{\psi} \psi} $
\beq
\delta Z_2 = - K_\ep  \,  Z_2 \, \delta \Gamma_B(\as^0,q^2)
\label{ZV:rec}
{},
\eeq
\beq
\delta Z_{\bar{\psi} \psi} = - K_e \,  Z_{\bar{\psi} \psi} \, \delta \Gamma^S_B(\as^0,q^2)
\label{ZS:rec}
{}.
\eeq
Eqs. (\ref{Zm:def},\ref{ZV:rec},\ref{ZS:rec}) directly express $Z_m$ through
massless propagator-type (that is dependent on one external momentum only)
Feynman integrals (FI), denoted as {\em p-integrals} below.

There exists currently no direct way to analytically evaluate five-loop
p-integrals.  However, for a given five-loop
p-integral we need to know only its {\em pole} part in $\ep$ in the
limit of $\ep \to 0$. The proper use of this fact can significantly
simplify our task. The corresponding method---the so-called Infrared
Rearrangement (IRR)---first suggested in \cite{Vladimirov:1979zm} and
elaborated further in \cite{Kazakov:1979ik,Chetyrkin:1980pr,Tarasov:1980au} 
allows to effectively decrease the number of
loops to be computed by one\footnote{With the price that the resulting
 p-integrals  with one loop less  should be evaluated up to and {\em
including} their constant part in the small $\ep$-expansion.}.  In its
initial version IRR was not really universal; it was not applicable in
some (though rather rare) cases of complicated FI's.  The problem was
solved by  elaborating a special technique of combined recursive subtraction of {\em both}
IR and UV divergences --- the $R^*$-operation
\cite{Chetyrkin:1984xa,Chetyrkin:1996ez}.
Formally, the   $R^*$-operation is defined as a product
\beq
R^*= R \, \tilde{R} = \tilde{R}\,  R
{}, 
\eeq
where $R$ stands for  the Bogoliubov-Parasiuk R-operation \cite{Bogoliubov:1957gp,BogoliubovShirkov198008} 
and $\tilde{R}$ refers to a kind of IR  R-operation which subtracts recursively all IR divergences 
from a given (Euclidean) FI.  A detailed discussion of  the $R-$, $\tilde{R}-$ and $R^*$-operations
and their interplay can be found in \cite{Chetyrkin:2017ppe,Batkovich:2014rka}.
The technique of  the $R^*$-operation succeeds in
expressing the UV counterterm of every $L$-loop FI in
terms of divergent and finite parts of some $(L-1)$-loop massless
propagators.

In our case $L=5$ and, using IRR, one arrives at  around $10^5$
four-loop p-integrals (all Feynman diagrams  have been generated 
with the use of   QGRAF \cite{Nogueira:1991ex}). 
These were  reduced to 28
four-loop master p-integrals, which are known analytically, including
their finite parts, from \cite{Baikov:2010hf,Lee:2011jt} as well as
numerically from \cite{Smirnov:2010hd}.

We have computed the necessary p-integrals with a special version of
reduction\footnote{Note, that very recently there has been  significant
progress in developing direct reduction algorithms (not using $1/D$
expansion) \cite{Lee:2012cn,Lee:2013mka,Ueda:2016yjmq,Herzog:2016qas}.} which is
based on evaluating sufficiently many terms of the $1/D$ expansion
\cite{Baikov:2005nv} of the corresponding coefficient functions
\cite{Baikov:1996rk}. The algorithm was implemented in a dedicated    
FORM \cite{Vermaseren:2000nd,Steinhauser:2015wqa} program.

\newcommand{\ZSS}{Z_{\bar{\psi} \psi}}

\newcommand{\ZS}{Z_{\bar{\psi} \psi}}

\section{Explicit formulas for $L$-loop $Z_2$ and $Z_{\bar{\psi} \psi}$\, via $(L-1)$
p-integrals}

Let us start from eq.~\re{Gscalar:def} by  rewriting 
it as  follows 
\beq
\ZSS = 1-   K_\ep \, R' 
  \delta \Gamma^S(\as,q^2)
\label{ZS:R:def}
{}, \eeq 
where $K_\epsilon \left\{f(\epsilon)\right\}$ is the singular
part of the Laurent expansion of $f(\epsilon)$ in $\epsilon$ near $\epsilon=0$
and $R'$ stands for the ``incomplete'' $R$ operation which, when applied to a
FI, subtracts only all its  UV {\em sub}divergences not touching
the UV divergence of the FI as whole (for a good introductory text about renormalization theory and
$R$-operation see \cite{Kennedy_R}).

From a purely formal point of view eq.~\re{ZS:R:def} is  equivalent to
\re{ZS:rec}, however  the first relation is significantly more flexible as it does
not change its form if, for instance, some (or, even, all) propagators in FI's
contributing to $\delta \Gamma^S(\as,q^2)$ are made massive. To avoid any
confusion: by making a propagator $G_i(p)$ massive we mean its multiplication
by a factor $\frac{p^2}{p^2 -m_i^2}$.  The factor smoothes the behavior of
the corresponding Feynman integrands in  the region of small momenta of order
 $m$. It does not contribute in the only integration region relevant
for the UV counterm of a given diagram as whole (modulo power suppressed terms of order $m_i^2/p_i^2$ and higher) namely the region   where {\em all} loop momenta
are large.  This is an obvious consequence of the following statement
\cite{Collins:1974da}: any UV counterterm for any logarithmically divergent FI
integral does not depend on momenta and masses.

As a result we can transform  eq. \re{ZS:R:def} to
\beq
\ZS = 1-   K_\ep \, R' 
  \delta \Gamma^S(\as,\unl{m}, q^2)
\label{ZS:R:masses}
{}, \eeq 
where $\unl{m}= m_1, \dots $. The choice of
masses in eq.~\re{ZS:R:masses} and $q^2$ is constrained by only one
requirement: there should be no IR divergences in its right hand side. The
condition of IR safety can removed completely if one lets the
$\tilde{R}$-operation  remove all IR divergences from the rhs of
\re{ZS:R:masses}.

To proceed we first ignore any IR divergences and choose the $m$-propagators (that
is those  which are  massive in \re{ZS:R:masses}) in such a way that the 
resulting FI's get simpler for calculation (the necessary IR subtractions (if any)
will be dealt with later).

A popular possibility is to make {\bf all} propagators massive with the help
of {\em one and the same} mass and nullify $q$.  The resulting completely
massive vacuum graphs are then computed with one or another version of IBP
reduction\footnote{ The method is actively used since long 
\cite{Misiak:1994zw,4loopbetaqcd,beta_den_comp}.
 Very recent impressive
  results obtained with the method can be found in
  \cite{Luthe:2016xec,Luthe:2017ttc}.}. The advantages of this ``maximally
massive'' approach are that (i) no IR singularities may appear, (ii) global
(unlike local) gauge symmetry stays untouched and, finally, the global
renormalization with the help of Z-factors is not very different from the
standard one.  There is also a price: to compute $\ZS$ on $L$-loop level one
should deal with $L$-loop massive vacuum graphs which are  significantly more
complicated for calculation than $(L-1)$-loop p-integrals.

We will employ a ``minimally massive'' way, that is the choice of $q=0$ and
$\unl{m} = m,0,0,0, \dots$ which means that only one line in every FI
contributing to $\Gamma^S$ is provided with a mass.

\ice{ (we will see later
  the formal reason why the ``obviously'' simplest choice of total
  nullification of all masses and $q$ is not a suitable one).
}

In principle, a particular choice of an m-propagator could be tuned
(diagramwise!) in many cases in such a way to avoid any IR
singularities\footnote{That was exactly the strategy used in the 
pioneering calculations within IRR method 
\cite{Vladimirov:1979ak,Kazakov:1979ik,Chetyrkin:1979bj,Tarasov:1980au,Gorishnii:1990vf}.} 
and, thus, to avoid any
use of $\tilde{R}$ operation.  However, such a tuning essentially prevents
any possibility to solve combinatoric of both $R$ and $\tilde{R}$ operations
globally in terms of corresponding Z-factors. This, in turn, forces us
either to perform an extremely tedious manual  diagram-wise renormalization
of UV/IR divergences or to  develop  highly non-trivial computer algebra
routines for its  automatization\footnote{Very recently such routines 
implementing  automatic UV and IR  renormalization on the level of separate diagrams
have been developed    \cite{Herzog:2017ohr}.}. 
\begin{figure}[tb]
  \begin{center}
\includegraphics[width=5cm]{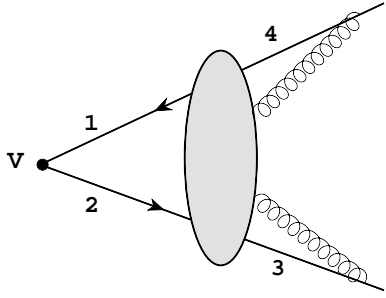}
\caption{\label{fig:1} All  distinguished fermion propagators available 
for a generic vector (and scalar)  vertex functon.  
      }
  \end{center}
\end{figure}

Every diagram contributing to $\Gamma^S$ has only four internal propagators
which could be specified globally without referencing to particular
graphs. (This is a necessary requirement allowing us to resolve combinatorics
of {\em both} UV and IR subtractions in terms of global Z-factors).  These
are shown on Fig. 1. Let us choose the fermion line  which flows into the
scalar current vertex (marked by 1 on Fig.~1) and make the corresponding propagator massive by
multiplying it by $p^2/(p^2-m^2)$. The result of applying  the $R'$
operation to the corresponding vertex function $\delta \Gamma^S(\as,m)$ can be
expressed  in terms of the bare vertex function and relevant Z-factors as 
\beq R'
\delta \Gamma^S(\as,m) = \delta \Gamma^S(\as^0,m) + \delta Z_{\bar{\psi}\psi} \,\delta \Gamma^S(\as^0,m=0)
\label{Rprime:GS:q:0}
{}.
\eeq
The rhs of the above equation is contaminated by  IR singularities; they appear 
not  only in the second term but also in the first one. 
The application of  $\tilde{R}$ operation to remove all IR divergent pieces from 
\re{Rprime:GS:q:0}  leads to:
\beq
\tilde{R}\, R'\delta \Gamma^S(\as,m) = \delta \Gamma^S(\as^0,m) +
\left(
 \delta \Gamma^S(\as^0,m)\,   + \
\delta Z_{\bar{\psi}\psi}\,
\right)
\cdot\,\fbox{$\displaystyle \left(\frac{1}{Z_{\bar{\psi}\psi}} - 1\right)$}
{}\, ,
\eeq
where we have boxed the IR Z-factor. The considerations for the vector case follow the same 
pattern with $Z_{\bar{\psi}\psi}$ replaced by $Z_V$.

Thus, the final formulas for $\delta Z_V$ and $\delta Z_{\bar{\psi}\psi}$ are:
\beq
\delta Z_2 = 
 -\delta \Gamma^V(\as^0,m)\, \frac{1}{Z_V}  + \frac{(Z_V-1)^2}{Z_V}
\label{ZV:fin}
{},
\eeq

\beq
\delta Z_{\bar{\psi}\psi} = 
 - \delta \Gamma^S(\as^0,m)\, \frac{1}{Z_{\bar{\psi}\psi}}  +  \frac{(Z_{\bar{\psi}\psi}-1)^2}{Z_{\bar{\psi}\psi}}
\label{ZS:fin}
{}.
\eeq

\section{Results}

Our result for the anomalous dimension

\begin{equation}
\gamma_m = - \sum_i (\gamma_m)_i a_s^{i+1}
{},
\end{equation}
reads (for completeness we also present  the  lower order 
terms \cite{Tarrach:1980up,Tarasov:1982gk,Larin:1993tq,Vermaseren:1997fq})
\begin{equation}
\gamma_m = - \sum_i (\gamma_m)_i a_s^{i+1}
\end{equation}
\beq
(\gamma_m)_0 = \frac{3}{4}\,C_F
{}, \hspace{10mm}
(\gamma_m)_1 \,=\, \frac{1}{4^2}\Biggl\{   
 \frac{3}{2}\,\cfB
-\frac{10}{3}\, \cfAtrAnlA
+
 \frac{97}{6}
\cfAcaA
\Biggr\}
\label{gmL1_2}
{},
\nonumber
\eeq

\begin{eqnarray}
(\gamma_m)_2 &=&
\frac{1}{4^3} \Biggl\{  
 \frac{129}{2}
\cfC
{-}
\left[
46 
-48  \sbz \zeta_{3}
\right]
\cfBtrAnlA
-\frac{140}{27}
\cfAtrBnlB
-\frac{129}{4}
\cfBcaA
\nonumber\\
&{-}&
\left[
\frac{556}{27} 
+48  \sbz \zeta_{3}
\right]
\cfAtrAnlAcaA
+\frac{11413}{108}
\cfAcaB
\Biggr\}
{},
\nonumber
\end{eqnarray}

\begin{eqnarray}
(\gamma_m)_3 &=&
\frac{1}{4^4} \Biggl\{ 
\cfD
\left[
-\frac{1261}{8} 
-336  \sbz \zeta_{3}
\right]
{+}\cfCtrAnlA
\left[
-\frac{280}{3} 
+552  \sbz \zeta_{3}
-480  \sbz \zeta_{5}
\right]
\nonumber\\
&{+}&\cfBtrBnlB
\left[
\frac{304}{27} 
-160  \sbz \zeta_{3}
+96  \sbz \zeta_{4}
\right]
{+}\cfAtrCnlC
\left[
-\frac{664}{81} 
+\frac{128}{9}  \sbz \zeta_{3}
\right]
\nonumber\\
&{+}&\cfCcaA
\left[
\frac{15349}{12} 
+316  \sbz \zeta_{3}
\right]
{+}\cfBtrAnlAcaA
\left[
-\frac{8819}{27} 
+368  \sbz \zeta_{3}
-264  \sbz \zeta_{4}
+80  \sbz \zeta_{5}
\right]
\nonumber\\
&{+}&\cfAtrBnlBcaA
\left[
\frac{1342}{81} 
+160  \sbz \zeta_{3}
-96  \sbz \zeta_{4}
\right]
{+}\cfBcaB
\left[
-\frac{34045}{36} 
-152  \sbz \zeta_{3}
+440  \sbz \zeta_{5}
\right]
\nonumber\\
&{+}&\cfAtrAnlAcaB
\left[
-\frac{65459}{162} 
-\frac{2684}{3}  \sbz \zeta_{3}
+264  \sbz \zeta_{4}
+400  \sbz \zeta_{5}
\right]
{+}\cfAcaC
\left[
\frac{70055}{72} 
+\frac{1418}{9}  \sbz \zeta_{3}
-440  \sbz \zeta_{5}
\right]
\nonumber\\
&{+}&\dFFdRinvAnlA
\left[
64 
-480  \sbz \zeta_{3}
\right]
{+}\dFARinvA
\left[
-32 
+240  \sbz \zeta_{3}
\right]
\Biggr\}
{},
\nonumber
\end{eqnarray}


\begin{eqnarray}
(\gamma_m)_4  &=& 
\frac{1}{4^5} \Biggl\{
\cfE
\left[
\frac{50995}{8} 
+848  \sbz \zeta_{3}
+2080  \sbz \zeta_{5}
\right]
\nonumber\\
&{+}&\cfDtrAnlA
\left[
-\frac{48797}{36} 
+6888  \sbz \zeta_{3}
-672  \sbz \zeta_{4}
-\frac{37000}{3}  \sbz \zeta_{5}
+6720  \sbz \zeta_{7}
\right]
\nonumber\\
&{+}&\cfCtrBnlB
\left[
\frac{45253}{54} 
-\frac{38416}{9}  \sbz \zeta_{3}
+896  \,\zeta_3^2
+1304  \sbz \zeta_{4}
+\frac{8000}{3}  \sbz \zeta_{5}
-1600  \sbz \zeta_6
\right]
\nonumber\\
&{+}&\cfBtrCnlC
\left[
\frac{8966}{81} 
+\frac{352}{3}  \sbz \zeta_{3}
-320  \sbz \zeta_{4}
+\frac{512}{3}  \sbz \zeta_{5}
\right]
\nonumber\\
&{+}&\cfDcaA
\left[
-\frac{2565029}{144} 
-13060  \sbz \zeta_{3}
+1848  \sbz \zeta_{4}
+\frac{29600}{3}  \sbz \zeta_{5}
\right]
\nnb
\eea
\bea
&{+}&\cfCtrAnlAcaA
\left[
-\frac{406861}{108} 
-\frac{18013}{9}  \sbz \zeta_{3}
+1760  \,\zeta_3^2
-2360  \sbz \zeta_{4}
+5620  \sbz \zeta_{5}
+4400  \sbz \zeta_6
-3360  \sbz \zeta_{7}
\right]
\nonumber\\
&{+}&\cfCcaB
\left[
\frac{2625197}{72} 
+24174  \sbz \zeta_{3}
-1738  \sbz \zeta_{4}
-\frac{99140}{3}  \sbz \zeta_{5}
-9240  \sbz \zeta_{7}
\right]
\nonumber\\
&{+}&\cfBtrBnlBcaA
\left[
\frac{375373}{486} 
+\frac{5996}{9}  \sbz \zeta_{3}
-\frac{6976}{3}  \,\zeta_3^2
+1864  \sbz \zeta_{4}
-\frac{12448}{9}  \sbz \zeta_{5}
+\frac{800}{3}  \sbz \zeta_6
\right]
\nonumber\\
&{+}&\cfBtrAnlAcaB
\left[
\frac{753557}{1944} 
+\frac{72194}{9}  \sbz \zeta_{3}
-\frac{5456}{3}  \,\zeta_3^2
-3226  \sbz \zeta_{4}
-\frac{81464}{9}  \sbz \zeta_{5}
+\frac{2200}{3}  \sbz \zeta_6
+560  \sbz \zeta_{7}
\right]
\nonumber\\
&{+}&\cfBcaC
\left[
-\frac{25256617}{972} 
-\frac{50642}{3}  \sbz \zeta_{3}
-\frac{1936}{3}  \,\zeta_3^2
+836  \sbz \zeta_{4}
+\frac{374180}{9}  \sbz \zeta_{5}
-\frac{12100}{3}  \sbz \zeta_6
+6160  \sbz \zeta_{7}
\right]
\nonumber\\
&{+}&\cfAtrDnlD
\left[
-\frac{1040}{81} 
-\frac{1280}{81}  \sbz \zeta_{3}
+\frac{256}{9}  \sbz \zeta_{4}
\right]
\nonumber\\
&{+}&\cfAtrCnlCcaA
\left[
\frac{18667}{243} 
+\frac{21472}{81}  \sbz \zeta_{3}
+\frac{2176}{9}  \sbz \zeta_{4}
-\frac{4096}{9}  \sbz \zeta_{5}
\right]
\nonumber\\
&{+}&\cfAtrBnlBcaB
\left[
\frac{27418}{243} 
+\frac{87722}{27}  \sbz \zeta_{3}
+\frac{4288}{3}  \,\zeta_3^2
-\frac{9364}{3}  \sbz \zeta_{4}
-\frac{2944}{3}  \sbz \zeta_{5}
+\frac{4000}{3}  \sbz \zeta_6
\right]
\nonumber\\
&{+}&\cfAtrAnlAcaC
\left[
-\frac{4994047}{972} 
-\frac{959759}{81}  \sbz \zeta_{3}
+\frac{176}{3}  \,\zeta_3^2
+\frac{54925}{9}  \sbz \zeta_{4}
+\frac{139234}{9}  \sbz \zeta_{5}
-\frac{15400}{3}  \sbz \zeta_6
-3920  \sbz \zeta_{7}
\right]
\nonumber\\
&{+}&\cfAcaD
\left[
\frac{22663417}{1944} 
+\frac{418936}{81}  \sbz \zeta_{3}
+\frac{1936}{3}  \,\zeta_3^2
-\frac{15697}{18}  \sbz \zeta_{4}
-\frac{64405}{3}  \sbz \zeta_{5}
+\frac{12100}{3}  \sbz \zeta_6
+3080  \sbz \zeta_{7}
\right]
\nonumber\\
&{+}&\dFFdRinvAnlAcfA
\left[
-352 
-4928  \sbz \zeta_{3}
+11840  \sbz \zeta_{5}
\right]
\nonumber\\
&{+}&\dFARinvAcfA
\left[
6768 
+3840  \sbz \zeta_{3}
-25440  \sbz \zeta_{5}
\right]
\nonumber\\
&{+}&\dFFdRinvAnlBtrA
\left[
-\frac{6896}{9} 
+\frac{7312}{3}  \sbz \zeta_{3}
-768  \sbz \zeta_{4}
-\frac{2240}{3}  \sbz \zeta_{5}
\right]
\nonumber\\
&{+}&\dFARinvAnlAtrA
\left[
\frac{3040}{9} 
-\frac{14920}{3}  \sbz \zeta_{3}
-1408  \,\zeta_3^2
-144  \sbz \zeta_{4}
+\frac{3520}{3}  \sbz \zeta_{5}
\right]
\nonumber\\
&{+}&\dFFdRinvAnlAcaA
\left[
\frac{25384}{9} 
-\frac{16720}{3}  \sbz \zeta_{3}
+1408  \,\zeta_3^2
+2640  \sbz \zeta_{4}
-\frac{35680}{3}  \sbz \zeta_{5}
\right]
\nonumber\\
&{+}&\dFARinvAcaA
\left[
-\frac{76784}{9} 
-\frac{1208}{3}  \sbz \zeta_{3}
-3872  \,\zeta_3^2
-1320  \sbz \zeta_{4}
+23200  \sbz \zeta_{5}
+1232  \sbz \zeta_{7}
\right]
\nonumber\\
&{+}&\dAARinvAtrA
\left[
358 
+\frac{4988}{3}  \sbz \zeta_{3}
+3872  \,\zeta_3^2
+264  \sbz \zeta_{4}
-\frac{15640}{3}  \sbz \zeta_{5}
-1232  \sbz \zeta_{7}
\right]
\Biggr\}
{}.
\label{gmL5}
\end{eqnarray}
Here $\zeta$ is the Riemann zeta-function (with
$\zeta_3=1.2020569\dots$, $\zeta_4=1.0823232\dots$,
$\zeta_5=1.0369278\dots$, $\zeta_6= 1.0173431\dots$,
$\zeta_7=1.0083493\dots$).  
$C_F$ and $ C_A$ are the quadratic Casimir
operators of the quark $[T^a T^a]_{ij} = C_F \delta_{ij}$ and the
adjoint $[C^a \, C^a]_{bd} = C_A \,  \delta_{bd}$, $(C^a)_{bc} = -if^{abc}$ 
representations of the Lie algebra. $n_f$ stands for  the number of
quark flavors, $d_R$ is dimension of the
quark repesenation of the gauge group and $T_f$ refers to the trace
normalization $\mathrm{tr}\,(T^a T^b) = T_f \,\delta^{ab}$. 
The  higher order group invariants are defined as
contraction between  the following (symmetrical) tensors
(see \cite{Vermaseren:1997fq,vanRitbergen:1998pn})
\begin{eqnarray}
 d_F^{a b c d} & = & \frac{1}{6 } {\rm Tr }  \left[
   T^a T^b T^c T^d
 + T^a T^b T^d T^c 
 + T^a T^c T^b T^d  \right. \nonumber \\
 & & \left. \hspace{4mm}
 + T^a T^c T^d T^b 
 + T^a T^d T^b T^c 
 + T^a T^d T^c T^b  \hspace{1mm}
 \right]
{},
\\
\label{eq:CCCC}
 d_A^{a b c d} & = & \frac{1}{6} {\rm Tr }  \left[
   C^a C^b C^c C^d
 + C^a C^b C^d C^c 
 + C^a C^c C^b C^d  \right. \nonumber \\
 & & \left. \hspace{4mm}
 + C^a C^c C^d C^b 
 + C^a C^d C^b C^c 
 + C^a C^d C^c C^b  \hspace{1mm}
 \right]
{}.
\end{eqnarray}
Note that for the gauge group $SU(3)$, after setting
$C_F=4/3$, $C_A=3, T_f =1/2$, 
\mbox{$d_F^{abcd}\,  d_F^{abcd} =5/6$},
\mbox{$d_F^{abcd}\,  d_A^{abcd} =15/2$},
\mbox{$d_A^{abcd}\,  d_A^{abcd} =135$}
\ice{
id  dFF3 = 5/6 ;
id  dFA4 = 15/2 ;
id  dFF4 = 5/12 ;
id  dAA4 = 135 ;
}
we exactly reproduce the SU(3) result for $\gamma_m$ given in
\cite{Baikov:2014qja}.

For the special case of QED, setting  $C_F=1$, $T_F=1$, $C_A=0, d_A^{abcd}=0, d_F^{abcd}
d_F^{abcd} =1$,   and $d_R=1$
we arrive at
\begin{eqnarray}
(\gamma^{QED}_m)_{0}&=&  \frac{3}{4} 
{},\hspace{10mm}
(\gamma^{QED}_m)_{1} = \frac{3}{32}-\frac{5}{24}\,n_f 
{},
\nonumber
\end{eqnarray}

\begin{eqnarray}
(\gamma^{QED}_m)_{2} &=& 
\frac{1}{4^3} \Biggl\{  
 \frac{129}{2}
{-} \,n_f 
\left[
46 
-48  \sbz \zeta_{3}
\right]
{-} \,\frac{140}{27}
n_f^2
\Biggr\}
{},
\nonumber
\end{eqnarray}

\begin{eqnarray}
(\gamma^{QED}_m)_{3} &=&  
\frac{1}{4^4} \Biggl\{  
{-}
\left[
\frac{1261}{8} 
+336  \sbz \zeta_{3}
\right]
{-} \,n_f 
\left[
\frac{88}{3} 
-72  \sbz \zeta_{3}
+480  \sbz \zeta_{5}
\right]
{+} \, n_f^2
\left[
\frac{304}{27} 
-160  \sbz \zeta_{3}
+96  \sbz \zeta_{4}
\right]
\nonumber\\
&{-}& \, n_f^3
\left[
\frac{664}{81} 
-\frac{128}{9}  \sbz \zeta_{3}
\right]
\Biggr\}
{},
\nonumber
\end{eqnarray}

\begin{eqnarray}
(\gamma^{QED}_m)_{4} &=& 
\frac{1}{4^5} \Biggl\{  
\left[
\frac{50995}{8} 
+848  \sbz \zeta_{3}
+2080  \sbz \zeta_{5}
\right]
{+} \,n_f 
\left[
-\frac{61469}{36} 
+1960  \sbz \zeta_{3}
-672  \sbz \zeta_{4}
-\frac{1480}{3}  \sbz \zeta_{5}
+6720  \sbz \zeta_{7}
\right]
\nonumber\\
&{+}& \, n_f^2
\left[
\frac{3877}{54} 
-\frac{16480}{9}  \sbz \zeta_{3}
+896  \,\zeta_3^2
+536  \sbz \zeta_{4}
+1920  \sbz \zeta_{5}
-1600  \sbz \zeta_6
\right]
\nonumber\\
&{+}& \, n_f^3
\left[
\frac{8966}{81} 
+\frac{352}{3}  \sbz \zeta_{3}
-320  \sbz \zeta_{4}
+\frac{512}{3}  \sbz \zeta_{5}
\right]
{+} \, n_f^4
\left[
-\frac{1040}{81} 
-\frac{1280}{81}  \sbz \zeta_{3}
+\frac{256}{9}  \sbz \zeta_{4}
\right]
\Biggr\}
{}.
\label{gmQEDL5}
\end{eqnarray}
\section{The  puzzle of  $\zeta_4$ and  $\zeta_6$}

\newcommand{\lQ}{\ell_Q}

In this Section we will consider  exclusively ``quantities'' expressible (read computable)
in terms of massless propagators (p-integrals). 
For any  such  quantity  computed up to some  number of loops  
the result always contains some quite limited number of the irrational constants.

More precisely, if a quantity, say,  $P(Q^2)$  can be expressed in terms of
p-integrals with the loop number $L$ {\em not exceeding four}, then 
the complete list of the  irrational constants reads
$\zeta_3, \zeta_4,\zeta_3^2,\zeta_3\zeta_4$, $\zeta_6$  and, finally,   $\zeta_7$. The reason is 
obvious, these and only these irrational constants  appear in all  4-loop master
p-integrals \cite{Baikov:2010hf}. If we consider $ L \le 3 $ then only  $\zeta_3, \zeta_4$ and $\zeta_5$
survive \cite{Chetyrkin:1980pr}. Finally, for $ L \le 2 $  there remains only  $\zeta_3$. 

It was already observed long ago that in practice  the real pattern of
appearance of the irrational constants is somewhat more limited than
the one  described above. 


Indeed, up to now the result of the explicit calculation of any
``physical'' (in the sense of having no anomalous dimension) quantity
$P$ has never contained even zetas, that is $\zeta_4$ and
$\zeta_6$.  Note that statement is applicable only to Euclidean
quantities; it is very well known that terms proportional $\pi^2$ are
routinely generated during the procedure of analytical continuation to the
Minkowskian  (negative) values of the momentum transfer $Q^2$.  

Two most prominent   examples are:  the Adler function $D(Q^2)$
and the  coefficient function  $C^{Bjp}$ in the Wilson expansion of  two vector currents
related to the  Bjorken sum rules from  polarized electron-nucleon scattering 
(both known up to oder $\alpha_s^4$ \cite{Baikov:2010je}).

Some reason  behind this remarkable pattern of absence of even zetas from
physical quantities $P(Q^2)$ was provided by an analysis of the transcendental
structure of the corresponding master p-integrals at  the  level of three
\cite{Broadhurst:1999xk} and four \cite{Baikov:2010hf} loops
correspondingly. In particular, in \cite{Baikov:2010hf} the following {\em theorem} has
been proven (by ``any'' p-integral in its formulation we mean arbitrary
p-integral with the number of loops not exceeding four):
\vspace{5mm}

\noindent {\bf Theorem}
\begin{enumerate}
\item Any  p-integral,  {\em finite} at $\ep \to 0 $,  does not contain even zetas
  $\{\zeta_{2n} \mid n \ge 2\}$ in the limit of $\ep \rightarrow 0$.

\item Any combination of p-integrals, {\em finite} at $\ep \to 0 $, like
\[
 \sum C_i(\ep) p_i {}, \ \ C_i = \sum_j  C_{ij} \ep^j
{},
\]
with the coefficient functions being  functions (not necessarily  finite at  $\ep \to  0$)   
with purely rational coefficients $C_{ij}$,  will  {\em not} contain
even zetas in the limit of $\ep \rightarrow 0$ (while  {\em odd}  zetas  
\mbox{  $\{\zeta_{2n+1} \mid n \ge 1\}$ } are expected and indeed  appear in general).

\item Let $F(\ep)$ be any renormalized (and, thus, finite in the limit
of $\ep \to 0$) combination of any p-integrals. The {\em sole} source of
possible even zetas  in $F(0)$ is   the appearance of  zetas (not necessarily even) in the
renormalization factors involved in carrying out the renormalization of F.

\end{enumerate}

The theorem is a direct consequence of the following observation:
all explicit results for  4-loop master p-integrals 
do depend on only the following {\bf three} combinations of zetas:
\beq
\hat{\zeta}_3 = \zeta_3+\frac{3 \ep}{2}{\zeta_4}-\frac{5 \ep^3}{2}{\zeta_6},
\,\, 
\hat{\zeta}_5=\zeta_5+\frac{5\ep}{2}\zeta_6 \ \ \  \mbox{and} \ \  \ \zeta_7.
\label{hat_zetas}
\eeq

The third point   of the  theorem provides the reason behind   the absence
of   even zetas in  the Adler function  at order $\alpha_s^3$
and in the coefficient function $C^{Bjp}$ at at order $\alpha_s^4$.
Indeed, both quantities are  

(i) a {\em finite} combination of four-loop p-integrals; 

(ii) the corresponding renormalizations require only   the  charge
coupling renormalization  at maximum  three-loop order which is free from any zetas.

The fact that the Adler function  at order $\alpha_s^4$ is also free from  even zetas is 
not explained  by the theorem. The reason is that in this order the Adler function
is directly expressible in terms of {\em five}-loop p-integrals.    The fact could  be understood 
if the five-loop master p-integrals would obey a property similar to  \re{hat_zetas}. 

The hypothesis (suggested in \cite{Baikov:2010hf}) is supported by another
5-loop physical quantity related to the scalar correlator. The
corresponding Adler function is defined as follows:
\begin{equation}
\widetilde{D}(Q^2) =  \frac{Q^2}{6}
\frac{\mathrm{d}}{\mathrm{d} Q^2} \frac{\widetilde{\Pi}(Q^2)}{Q^2}
=
\int_0^\infty \frac{ Q^2\ \widetilde{R}(s) d s }{(s+Q^2)^2}
{}\ ,
\label{Adler_scalar}
\end{equation}
\beq
\nonumber
 \widetilde{D}(Q^2) = 1+ \sum_{i=1}^{\infty} \  \widetilde{d}_i(\lQ)\, a_s^{i}(\mu)
{},
\eeq
where  $\mu$ is  the $\msbar$ renormalization scale 
and $\lQ = \ell_{ Q^2/\mu^2} $. The available results
for $\tilde{d}_i$ read     \cite{Baikov:2005rw} (note 
that in order to save  some space we set the number of quark flavours $n_f=3$
everywhere in this Section):

\beq
{\tilde{d}_1 =  } 
 \frac{17}{3}
-2 \,l_{Q}
{},
\,\, 
{\tilde{d}_2 =  } 
\frac{9631}{144} 
-\frac{35}{2}  \,\zeta_{3} 
-\frac{95}{3}\,\lQ
+
 \frac{17}{4} \,l_Q^2
{},
\label{scalarD2}
\eeq
\begin{eqnarray}
\tilde{d}_3 &=&   
\frac{4748953}{5184} 
-\frac{91519}{216}  \,\zeta_{3}
-\frac{5}{2}  \sbz \zeta_{4}
+\frac{715}{12}  \sbz \zeta_{5}
\nnb
\\
&+& \,l_{Q}
\left[
-\frac{4781}{9} 
+\frac{475}{4}  \,\zeta_{3}
\right]
{+} 
 \frac{229}{2}
\,l_Q^2
-\frac{221}{24}
\,l_Q^3
{},
\label{scalarD3}
\end{eqnarray}
\begin{eqnarray}
\tilde{d}_4 &=&   
\frac{7055935615}{497664} 
-\frac{46217501}{5184}  \,\zeta_{3}
+\frac{192155}{216}  \,\zeta_3^2
-\frac{17455}{576}  \sbz \zeta_{4}
\nnb
\\
&{+}&\frac{455725}{432}  \sbz \zeta_{5}
-\frac{625}{48}  \,\zeta_{6}
-\frac{52255}{256}  \,\zeta_{7}
\nonumber\\
&{+}& \,l_{Q}
\left[
-\frac{97804997}{10368} 
+\frac{1166815}{288}  \,\zeta_{3}
+5  \sbz \zeta_{4}
-\frac{24025}{48}  \sbz \zeta_{5}
\right]
\nonumber\\
&{+}& \,l_Q^2
\left[
\frac{3008729}{1152} 
-\frac{16785}{32}  \,\zeta_{3}
\right]
+ \,l_Q^3
\left[
-\frac{51269}{144}\right]
{+} \,l_Q^4
\left[
 \frac{7735}{384}\right]
{}.
\label{scalarD4}
\end{eqnarray}
Unlike the vector  case the scalar Adler function is full of even zetas starting already from 
three loops (the coefficient $\tilde{d}_3$). This  is a natural consequence of the two facts

\noindent 
(i)
the quantity by itself is  {\em not} scale  invariant but meets the following evolution equation
\beq
\mu^2 \frac{ {\mathrm d} }{ {\mathrm d} \mu^2}\,  \widetilde{D}(\mu^2/Q^2, \alpha_s(\mu)) = -2\, \g_m
\label{RG:wD1}
\eeq
or, equivalently, 
\beq
\mu^2 \frac{ \prd }{\prd \mu^2}\,  \widetilde{D}(\mu^2/Q^2, \alpha_s(\mu)) =
 -\left(2\, \g_m + \beta(\as)\,  \frac{ \prd }{\prd \as}\right) \widetilde{D}
\label{RG:wD2}
{}. 
\eeq 
(ii) the quark mass anomalous dimension starts to  depend on $\zeta_3$ 
at 3 loops and on $\zeta_4$ at 4 loops. 
\newcommand{\lmuQ}{\ell_{\mu Q}}

Let us now consider a related quantity, namely, 
\beq
\tilde{S}(\mu^2/Q^2, \alpha_s(\mu)) \equiv Q^2 \frac{ \prd }{\prd Q^2}\, \ln \widetilde{D}
{},
\eeq
which, obviously, is scale-invariant and meets the evolution equation:
\beq
\mu^2 \frac{ {\mathrm d} }{ {\mathrm d} \mu^2}\,  \tilde{S}(\mu^2/Q^2, \alpha_s(\mu)) =0
{}.
\eeq
Direct calculation gives:
\beq
\tilde{S}= \sum_{i=1,\infty } \tilde{s}_{i}(\lmuQ) \, \as^i(\mu)
\label{S:def}
{},
\eeq
with $\lmuQ = \ln \frac{\mu^2}{Q^2}$ and
\beq
\tilde{s}_1 = -2, \ 
{\tilde{s}_2 =  } 
-\frac{61}{3}
{+} 
 \frac{9}{2}\,l_{Q}
{},
\ \ 
\nnb
\tilde{s}_3 =  
-\frac{20321}{72} 
+\frac{335}{4}  \,\zeta_{3}
{+} \,l_{Q}
\left[
 \frac{199}{2}\right]
-\frac{81}{8}\,l_Q^2
{},
\ \ 
\label{S3}
\eeq
\begin{eqnarray}
\tilde{s}_4 =  
&{-}&
\frac{48127465}{10368} 
+\frac{2050813}{864}  \,\zeta_{3}
-\frac{18305}{48}  \sbz \zeta_{5}
\nnb
\\
&{+}& \,l_{Q}
\left[
\frac{400873}{192} 
-\frac{9045}{16}  \,\zeta_{3}
\right]
{+} 
-\frac{5661}{16} \,l_Q^2
{+} 
 \frac{729}{32}\,l_Q^3
{}.
\label{S4}
\end{eqnarray}
We observe that within the physical quantity S all even zetas abounding in
$\widetilde{D}$ neatly cancel each other not only in orders $\as^2,\ \as^3$
(as guaranteed by the theorem) but in  \mbox{order $\alpha_s^4$}, too.

In fact, the absence of the even zetas in $S$ at order $\as^4$ was
discovered in \cite{Vermaseren:1997fq} ten years {\em before} the result for
$\tilde{d}_4$ became available. Indeed, as $S$ depends on $\mu$ and $Q^2$ only
via the combination $\mu^2/Q^2$,  the evolution equation $\label{RG:wD1}$ allows 
to {\em construct } the derivative $ \frac{ \prd }{\prd Q^2}\, \widetilde{D}$
at $(L+1)$-loop level merely from the knowledge of $\widetilde{D}$ at $L$ loops,
the $\g_m$ and the $\beta$-function at $(L+1)$ loops.

As a result the expression \re{S5} for $S$ was constructed\footnote{
To be honest, the authors of this work were dealing not with  the function $S$
but with a closely related one; the difference is not essential for  our reasonings.}
 by the authors of
\cite{Vermaseren:1997fq} from  $ \widetilde{D}$ at  three loops and the  quark mass
anomalous dimension  at four loops  (the latter was the main result of their publication). In
fact, the absence of $\zeta_4$ in $S$ was considered there as an extra
cross-check for  the both calculations of $\gamma_m$ and $S$.

Now, with the 5-loop $\g_m$ at hands we can  easily construct the function $S$ at
 six-loop level (that is to  order $\alpha_s^5$!) The result
for $S_{5}$ read
\begin{eqnarray}
\tilde{s}_5 =   
&{}& \hspace{-4mm}
-\frac{43177218695}{497664} 
+\frac{3589509737}{62208}  \,\zeta_{3}
-\frac{20395}{3}  \,\zeta_3^2
-\frac{1335}{128}  \sbz \zeta_{4}
-\frac{200406415}{31104}  \sbz \zeta_{5}
+\frac{3285415}{2304}  \,\zeta_{7}
\nonumber\\
&{+}& \,l_{Q}
\left[
\frac{105141365}{2304} 
-\frac{2144623}{96}  \,\zeta_{3}
+\frac{54915}{16}  \sbz \zeta_{5}
\right]
{+} \,l_Q^2
\left[
-\frac{2569659}{256} 
+\frac{81405}{32}  \,\zeta_{3}
\right]
\nonumber\\
&{+}& \,l_Q^3
\left[
 \frac{17631}{16}\right]
{+} \,l_Q^4
\left[
-\frac{6561}{128}\right]
{}
\label{S5}
\end{eqnarray}
and it  indeed  contains $\zeta_4$.


\section{Conclusions}

We have presented the result for  the five-loop quark mass anomalous
dimension $\g_m$ for the  case of a generic gauge group as well as explicit formulas
which we have used. We have demonstrated that that at the level  $\alpha_s^5$
the even zetas do appear in physical observables  expressible in terms of 
massless propagators.

The work by K.~G.~Chetykin and J.~H.~K\"uhn was supported by the Deutsche
Forschungsgemeinschaft through CH1479/1-1.  The work of P.~A.~Baikov was 
supported in part by grant NSh-7989.2016.2 of the President of Russian
Federation.

Note added: the calculation of $\g_m$ presented here was about its completion
when we were informed  about the existence of a result
\cite{Luthe:2016xec} (not then yet submitted to the Archive) for the same
quantity. After  a few days we have got  our result \re{gmL5} which happens
to be  in full agreement with that from \cite{Luthe:2016xec}.
We thank  the authors of \cite{Luthe:2016xec} for informing us on their
results.


\providecommand{\href}[2]{#2}\begingroup\raggedright\endgroup

\ed

\mbib
\begin{thebibliography}{10}

\bibitem{Baikov:2014qja}
P.~A. Baikov, K.~G. Chetyrkin and J.~H. K{\"u}hn, \emph{{Quark Mass and Field
  Anomalous Dimensions to ${\cal O}(\alpha_s^5)$}},
  \href{http://dx.doi.org/10.1007/JHEP10(2014)076}{\emph{JHEP} {\bf 10} (2014)
  76}, [\href{https://arxiv.org/abs/1402.6611}{{\tt 1402.6611}}].

\bibitem{Tarrach:1980up}
R.~Tarrach, \emph{The pole mass in perturbative qcd}, {\emph{Nucl. Phys.} {\bf
  B183} (1981) 384}.

\bibitem{Tarasov:1982gk}
O.~Tarasov, \emph{{Anomalous Dimensions Of Quark Masses In Three Loop
  Approximation}}, .

\bibitem{Larin:1993tq}
S.~Larin, \emph{{The Renormalization of the axial anomaly in dimensional
  regularization}},
  \href{http://dx.doi.org/10.1016/0370-2693(93)90053-K}{\emph{Phys.Lett.} {\bf
  B303} (1993) 113--118}, [\href{https://arxiv.org/abs/hep-ph/9302240}{{\tt
  hep-ph/9302240}}].

\bibitem{Chetyrkin:1997dh_good}
K.~G. Chetyrkin, \emph{{Quark mass anomalous dimension to ${\cal
  O}(\alpha_s^4)$}},
  \href{http://dx.doi.org/10.1016/S0370-2693(97)00535-2}{\emph{Phys. Lett.}
  {\bf B404} (1997) 161--165},
  [\href{https://arxiv.org/abs/hep-ph/9703278}{{\tt hep-ph/9703278}}].

\bibitem{Vermaseren:1997fq}
J.~A.~M. Vermaseren, S.~A. Larin and T.~van Ritbergen, \emph{{The 4-loop quark
  mass anomalous dimension and the invariant quark mass}}, {\emph{Phys. Lett.}
  {\bf B405} (1997) 327--333},
  [\href{https://arxiv.org/abs/hep-ph/9703284}{{\tt hep-ph/9703284}}].

\bibitem{Baikov:2014pja}
K.~Chetyrkin, P.~Balkov and J.~K{\"u}hn, \emph{{Towards QCD running in 5 loops:
  quark mass anomalous dimension}}, {\emph{PoS} {\bf RADCOR2013} (2013) 056},
  [\href{https://arxiv.org/abs/1402.6606}{{\tt 1402.6606}}].

\bibitem{Baikov:2016tgj}
P.~A. Baikov, K.~G. Chetyrkin and J.~H. K{\"u}hn, \emph{{Five-Loop Running of
  the QCD coupling constant}},  \href{https://arxiv.org/abs/1606.08659}{{\tt
  1606.08659}}.

\bibitem{Chetyrkin:2016uhw}
K.~Chetyrkin, P.~Baikov and J.~K{\"u}hn, \emph{{The $\beta$-function of Quantum
  Chromodynamics and the effective Higgs-gluon-gluon coupling in five-loop
  order}}, {\emph{PoS} {\bf LL2016} (2016) 010}.

\bibitem{Chetyrkin:2016pbz}
K.~Chetyrkin and J.~K{\"u}hn, \emph{{Precision Measurements in
  Electron-Positron Annihilation: Theory and Experiment}}, {\emph{PoS} {\bf
  LL2016} (2016) 047}.

\bibitem{Baikov2016}
P.~A. Baikov, K.~G. Chetyrkin and J.~H. K{\"u}hn, \emph{{Five-Loop Results in
  Perturbative QCD}},  \href{https://arxiv.org/abs/1402.6611}{{\tt 1402.6611}}.

\bibitem{tHooft:1972fi}
G.~'t~Hooft and M.~J.~G. Veltman, \emph{{Regularization and Renormalization of
  Gauge Fields}},
  \href{http://dx.doi.org/10.1016/0550-3213(72)90279-9}{\emph{Nucl. Phys.} {\bf
  B44} (1972) 189--213}.

\bibitem{Bardeen:1978yd}
W.~A. Bardeen, A.~J. Buras, D.~W. Duke and T.~Muta, \emph{Deep inelastic
  scattering beyond the leading order in asymptotically free gauge theories},
  {\emph{Phys. Rev.} {\bf D18} (1978) 3998}.

\bibitem{Chetyrkin:1996ia}
K.~G. Chetyrkin, J.~H. K{\"u}hn and A.~Kwiatkowski, \emph{{QCD corrections to
  the $e^{+} e^{-}$ cross-section and the $Z$ boson decay rate: Concepts and
  results}}, \href{http://dx.doi.org/10.1016/S0370-1573(96)00012-9}{\emph{Phys.
  Rept.} {\bf 277} (1996) 189--281}.

\bibitem{Vladimirov:1979zm}
A.~A. Vladimirov, \emph{{Method For Computing Renormalization Group Functions
  In Dimensional Renormalization Scheme}},
  \href{http://dx.doi.org/10.1007/BF01018394}{\emph{Theor. Math. Phys.} {\bf
  43} (1980) 417}.

\bibitem{Kazakov:1979ik}
D.~I. Kazakov, O.~V. Tarasov and A.~A. Vladimirov, \emph{{Calculation of
  Critical Exponents by Quantum Field Theory Methods}}, {\emph{Sov. Phys. JETP}
  {\bf 50} (1979) 521}.

\bibitem{Chetyrkin:1980pr}
K.~G. Chetyrkin, A.~L. Kataev and F.~V. Tkachov, \emph{{New Approach to
  Evaluation of Multiloop Feynman Integrals: The Gegenbauer Polynomial x Space
  Technique}},
  \href{http://dx.doi.org/10.1016/0550-3213(80)90289-8}{\emph{Nucl. Phys.} {\bf
  B174} (1980) 345--377}.

\bibitem{Tarasov:1980au}
O.~V. Tarasov, A.~A. Vladimirov and A.~Y. Zharkov, \emph{The gell-mann-low
  function of qcd in the three loop approximation}, {\emph{Phys. Lett.} {\bf
  B93} (1980) 429--432}.

\bibitem{Chetyrkin:1984xa}
K.~G. Chetyrkin and V.~A. Smirnov, \emph{{$R^*$ Operation Corrected}},
  \href{http://dx.doi.org/10.1016/0370-2693(84)91291-7}{\emph{Phys. Lett.} {\bf
  B144} (1984) 419--424}.

\bibitem{Chetyrkin:1996ez}
K.~G. Chetyrkin, \emph{Corrections of order { $\alpha_s^3$} to { $R_{\rm had}$}
  in { pQCD} with light gluinos},
  \href{http://dx.doi.org/10.1016/S0370-2693(96)01478-5}{\emph{Phys. Lett.}
  {\bf B391} (1997) 402--412},
  [\href{https://arxiv.org/abs/hep-ph/9608480}{{\tt hep-ph/9608480}}].

\bibitem{Bogoliubov:1957gp}
N.~N. Bogoliubov and O.~S. Parasiuk, \emph{{On the Multiplication of the causal
  function in the quantum theory of fields}},
  \href{http://dx.doi.org/10.1007/BF02392399}{\emph{Acta Math.} {\bf 97} (1957)
  227--266}.

\bibitem{BogoliubovShirkov198008}
N.~Bogoliubov and D.~Shirkov, \emph{Introduction to the Theory of Quantized
  Fields}.
\newblock John Wiley {\&}Sons Inc, 3rd~ed., 8, 1980.

\bibitem{Chetyrkin:2017ppe}
K.~G. Chetyrkin, \emph{{Combinatorics of $\mathbf{R}$-, $\mathbf{R^{-1}}$-, and
  $\mathbf{R^*}$-operations and asymptotic expansions of feynman integrals in
  the limit of large momenta and masses}},
  \href{https://arxiv.org/abs/1701.08627}{{\tt 1701.08627}}.

\bibitem{Batkovich:2014rka}
D.~V. Batkovich and M.~Kompaniets, \emph{{Toolbox for multiloop Feynman
  diagrams calculations using $R^{*}$ operation}},
  \href{http://dx.doi.org/10.1088/1742-6596/608/1/012068}{\emph{J. Phys. Conf.
  Ser.} {\bf 608} (2015) 012068}, [\href{https://arxiv.org/abs/1411.2618}{{\tt
  1411.2618}}].

\bibitem{Nogueira:1991ex}
P.~Nogueira, \emph{Automatic feynman graph generation}, {\emph{J. Comput.
  Phys.} {\bf 105} (1993) 279--289}.

\bibitem{Baikov:2010hf}
P.~A. Baikov and K.~G. Chetyrkin, \emph{{Four-Loop Massless Propagators: an
  Algebraic Evaluation of All Master Integrals}},
  \href{http://dx.doi.org/10.1016/j.nuclphysb.2010.05.004}{\emph{Nucl. Phys.}
  {\bf B837} (2010) 186--220}, [\href{https://arxiv.org/abs/1004.1153}{{\tt
  1004.1153}}].

\bibitem{Lee:2011jt}
R.~N. Lee, A.~V. Smirnov and V.~A. Smirnov, \emph{{Master Integrals for
  Four-Loop Massless Propagators up to Transcendentality Weight Twelve}},
  \href{http://dx.doi.org/10.1016/j.nuclphysb.2011.11.005}{\emph{Nucl. Phys.}
  {\bf B856} (2012) 95--110}, [\href{https://arxiv.org/abs/1108.0732}{{\tt
  1108.0732}}].

\bibitem{Smirnov:2010hd}
A.~V. Smirnov and M.~Tentyukov, \emph{{Four-Loop Massless Propagators: a
  Numerical Evaluation of All Master Integrals}},
  \href{http://dx.doi.org/10.1016/j.nuclphysb.2010.04.020}{\emph{Nucl. Phys.}
  {\bf B837} (2010) 40--49}, [\href{https://arxiv.org/abs/1004.1149}{{\tt
  1004.1149}}].

\bibitem{Lee:2012cn}
R.~N. Lee, \emph{{Presenting LiteRed: a tool for the Loop InTEgrals
  REDuction}},  \href{https://arxiv.org/abs/1212.2685}{{\tt 1212.2685}}.

\bibitem{Lee:2013mka}
R.~N. Lee, \emph{{LiteRed 1.4: a powerful tool for reduction of multiloop
  integrals}}, \href{http://dx.doi.org/10.1088/1742-6596/523/1/012059}{\emph{J.
  Phys. Conf. Ser.} {\bf 523} (2014) 012059},
  [\href{https://arxiv.org/abs/1310.1145}{{\tt 1310.1145}}].

\bibitem{Ueda:2016yjmq}
T.~Ueda, B.~Ruijl and J.~A.~M. Vermaseren, \emph{{Forcer: a FORM program for
  4-loop massless propagators}}, {\emph{PoS} {\bf LL2016} (2016) 070},
  [\href{https://arxiv.org/abs/1607.07318}{{\tt 1607.07318}}].

\bibitem{Herzog:2016qas}
F.~Herzog, B.~Ruijl, T.~Ueda, J.~A.~M. Vermaseren and A.~Vogt, \emph{{FORM,
  Diagrams and Topologies}}, {\emph{PoS} {\bf LL2016} (2016) 073},
  [\href{https://arxiv.org/abs/1608.01834}{{\tt 1608.01834}}].

\bibitem{Baikov:2005nv}
P.~A. Baikov, \emph{A practical criterion of irreducibility of multi-loop
  feynman integrals}, {\emph{Phys. Lett.} {\bf B634} (2006) 325--329},
  [\href{https://arxiv.org/abs/hep-ph/0507053}{{\tt hep-ph/0507053}}].

\bibitem{Baikov:1996rk}
P.~A. Baikov, \emph{{Explicit solutions of the 3--loop vacuum integral
  recurrence relations}},
  \href{http://dx.doi.org/10.1016/0370-2693(96)00835-0}{\emph{Phys. Lett.} {\bf
  B385} (1996) 404--410}, [\href{https://arxiv.org/abs/hep-ph/9603267}{{\tt
  hep-ph/9603267}}].

\bibitem{Vermaseren:2000nd}
J.~A.~M. Vermaseren, \emph{{New features of FORM}},
  \href{https://arxiv.org/abs/math-ph/0010025}{{\tt math-ph/0010025}}.

\bibitem{Steinhauser:2015wqa}
M.~Steinhauser, T.~Ueda and J.~A.~M. Vermaseren, \emph{{Parallel versions of
  FORM and more}},
  \href{http://dx.doi.org/10.1016/j.nuclphysbps.2015.03.006}{\emph{Nucl. Part.
  Phys. Proc.} {\bf 261-262} (2015) 45--57},
  [\href{https://arxiv.org/abs/1501.07119}{{\tt 1501.07119}}].

\bibitem{Kennedy_R}
W.~E. Caswell and A.~D. Kennedy, \emph{Simple approach to renormalization
  theory}, \href{http://dx.doi.org/10.1103/PhysRevD.25.392}{\emph{Phys. Rev. D}
  {\bf 25} (1982) 392--408}.

\bibitem{Collins:1974da}
J.~C. Collins, \emph{{Normal Products in Dimensional Regularization}},
  \href{http://dx.doi.org/10.1016/S0550-3213(75)80010-1}{\emph{Nucl. Phys.}
  {\bf B92} (1975) 477}.

\bibitem{Misiak:1994zw}
M.~Misiak and M.~M{\"u}nz, \emph{{Two loop mixing of dimension five flavor
  changing operators}},
  \href{http://dx.doi.org/10.1016/0370-2693(94)01553-O}{\emph{Phys. Lett.} {\bf
  B344} (1995) 308--318}, [\href{https://arxiv.org/abs/hep-ph/9409454}{{\tt
  hep-ph/9409454}}].

\bibitem{4loopbetaqcd}
T.~van Ritbergen, J.~Vermaseren and S.~Larin, \emph{{The Four loop beta
  function in quantum chromodynamics}},
  \href{http://dx.doi.org/10.1016/S0370-2693(97)00370-5}{\emph{Phys. Lett.}
  {\bf B400} (1997) 379--384},
  [\href{https://arxiv.org/abs/hep-ph/9701390}{{\tt hep-ph/9701390}}].

\bibitem{beta_den_comp}
K.~G. Chetyrkin, M.~Misiak and M.~M{\"u}nz, \emph{{Beta functions and anomalous
  dimensions up to three loops}},
  \href{http://dx.doi.org/10.1016/S0550-3213(98)00122-9}{\emph{Nucl. Phys.}
  {\bf B518} (1998) 473--494},
  [\href{https://arxiv.org/abs/hep-ph/9711266}{{\tt hep-ph/9711266}}].

\bibitem{Luthe:2016xec}
T.~Luthe, A.~Maier, P.~Marquard and Y.~Schroder, \emph{{Five-loop quark mass
  and field anomalous dimensions for a general gauge group}},
  \href{https://arxiv.org/abs/1612.05512}{{\tt 1612.05512}}.

\bibitem{Luthe:2017ttc}
T.~Luthe, A.~Maier, P.~Marquard and Y.~Schroder, \emph{{Complete
  renormalization of QCD at five loops}},
  \href{https://arxiv.org/abs/1701.07068}{{\tt 1701.07068}}.

\bibitem{Vladimirov:1979ak}
A.~A. Vladimirov, \emph{Methods of multiloop calculations and the
  renormalization group analysis of $\phi^4$ theory}, {\emph{Theor. Math.
  Phys.} {\bf 36} (1979) 732}.

\bibitem{Chetyrkin:1979bj}
K.~G. Chetyrkin, A.~L. Kataev and F.~V. Tkachov, \emph{{Higher Order
  Corrections to {$\sigma_{tot}(e^+ e^- \to \mbox{Hadrons})$} in Quantum
  Chromodynamics}},
  \href{http://dx.doi.org/10.1016/0370-2693(79)90596-3}{\emph{Phys. Lett.} {\bf
  B85} (1979) 277}.

\bibitem{Gorishnii:1990vf}
S.~G. Gorishnii, A.~L. Kataev and S.~A. Larin, \emph{{The
  $O(\alpha^{3}_{s})$-corrections to $\sigma_{tot}(e^{+}e^{-}\rightarrow
  hadrons)$ and $\Gamma(\tau^{-} \rightarrow \nu_{\tau} + hadrons)$ in QCD}},
  \href{http://dx.doi.org/10.1016/0370-2693(91)90149-K}{\emph{Phys. Lett.} {\bf
  B259} (1991) 144--150}.

\bibitem{Herzog:2017ohr}
F.~Herzog, B.~Ruijl, T.~Ueda, J.~A.~M. Vermaseren and A.~Vogt, \emph{{The
  five-loop beta function of Yang-Mills theory with fermions}},
  \href{https://arxiv.org/abs/1701.01404}{{\tt 1701.01404}}.

\bibitem{vanRitbergen:1998pn}
T.~van Ritbergen, A.~N. Schellekens and J.~A.~M. Vermaseren, \emph{Group theory
  factors for feynman diagrams}, {\emph{Int. J. Mod. Phys.} {\bf A14} (1999)
  41--96}, [\href{https://arxiv.org/abs/hep-ph/9802376}{{\tt hep-ph/9802376}}].

\bibitem{Baikov:2010je}
P.~A. Baikov, K.~G. Chetyrkin and J.~H. K{\"u}hn, \emph{{Adler Function,
  Bjorken Sum Rule, and the Crewther Relation to Order $\alpha_s^4$ in a
  General Gauge Theory}},
  \href{http://dx.doi.org/10.1103/PhysRevLett.104.132004}{\emph{Phys. Rev.
  Lett.} {\bf 104} (2010) 132004}, [\href{https://arxiv.org/abs/1001.3606}{{\tt
  1001.3606}}].

\bibitem{Broadhurst:1999xk}
D.~J. Broadhurst, \emph{{Dimensionally continued multiloop gauge theory}},
  \href{https://arxiv.org/abs/hep-th/9909185}{{\tt hep-th/9909185}}.

\bibitem{Baikov:2005rw}
P.~A. Baikov, K.~G. Chetyrkin and J.~H. K{\"u}hn, \emph{{Scalar correlator at
  {${\cal O}(\alpha_s^4)$}, Higgs decay into b- quarks and bounds on the light
  quark masses}},
  \href{http://dx.doi.org/10.1103/PhysRevLett.96.012003}{\emph{Phys. Rev.
  Lett.} {\bf 96} (2006) 012003},
  [\href{https://arxiv.org/abs/hep-ph/0511063}{{\tt hep-ph/0511063}}].

\end{thebibliography}
